\documentclass[pra,
                twocolumn,
                preprintnumbers,
                amsmath,
                amssymb,
                showpacs,
                superscriptaddress]{revtex4-1}
\usepackage{graphicx} 
\usepackage{dcolumn}  
\usepackage{bm}       
\usepackage{amsfonts}
\usepackage{dsfont}
\usepackage{mathtools}
\usepackage[colorlinks=true,linkcolor=blue,citecolor=red,urlcolor=blue]{hyperref}
\usepackage{color}
\usepackage{natbib}
\newcommand{\eq}[1]{Eq.~(\ref{#1})}

\begin{document}

\title{Attosecond delay of xenon $4d$ photoionization at the giant resonance and Cooper minimum}

\author{Maia Magrakvelidze}
\altaffiliation{%
Current address: Department of Physics, Kansas State University, Manhattan, Kansas 66502, USA}
\affiliation{%
Department of Natural Sciences, D.L.\ Hubbard Center for Innovation and Entrepreneurship,
Northwest Missouri State University, Maryville, Missouri 64468, USA}

\author{Mohamed El-Amine Madjet}
\affiliation{%
Qatar Environment and Energy Research Institute, Hamad Bin Khalifa University, Qatar Foundation, P.O. Box 5825, Doha, Qatar}

\author{Himadri S. Chakraborty}
\email[]{himadri@nwmissouri.edu}
\affiliation{%
Department of Natural Sciences, D.L.\ Hubbard Center for Innovation and Entrepreneurship,
Northwest Missouri State University, Maryville, Missouri 64468, USA}

\date{\today}



\begin{abstract}
A Kohn-Sham time-dependent local-density-functional scheme is utilized to predict attosecond time delays of xenon $4d$ photoionization that involves the $4d$ giant dipole resonance and Cooper minimum. The fundamental effect of electron correlations to uniquely determine the delay at both regions is demonstrated. In particular, for the giant dipole resonance, the delay underpins strong collective effect, emulating the recent prediction at C$_{60}$ giant plasmon resonance [T. Barillot {\em et al}, Phys.\ Rev.\ A {\bf 91}, 033413 (2015)]. For the Cooper minimum, a qualitative similarity with a photorecombination experiment near argon $3p$ minimum [S. B. Schoun {\em et al.}, Phys.\ Rev.\ Lett. {\bf 112}, 153001 (2014)] is found. The result should encourage attosecond measurements of Xe $4d$ photoemission.
\end{abstract}

\maketitle 

\section{Introduction}

Access to photoabsorption dynamics in real time {\em via} coherent pump-probe experiments is the key to fundamentally explore electron correlation processes in matters. Recent interests of such experiments~\cite{schultze2010delay, neppl2012attosecond, cavalieri2007attosecond, klunder2011probing, guenot2012photoemission} owe to the technology in generating attosecond single pulse~\cite{hentschel, goulielmakis1} and pulse trains~\cite{paul2001observation, mairesse2003attosecond}. Of particular attraction are the experiments based on the interferometric metrology, namely, the reconstruction of attosecond beating by interference of two-photon transitions (RABITT)~\cite{paul2001observation}, in which photoelectrons emitted by a coherent XUV comb of odd harmonics (pump) driven by a tunable fundamental field subsequently absorb or emit a synchronized IR photon (probe). This produces even harmonic sidebands in the spectrogram which oscillate as a function of ;the pump-probe offset-time. The ionization time delay is then obtained by the ratio of the difference of the measured phases at consecutive sidebands to the sideband separation. Since the additional delay introduced by the IR probe pulse via the so-called Coulomb-laser coupling can be estimated independently and subtracted from the measured result~\cite{ivanov2011accurate, nagele2012time, dahlstrom2012theory}, this phase-energy difference approach in RABITT comensurates with the Wigner-Smith route~\cite{wigner1955lower, smith1960lifetime} to determine emission time delay that involves energy-differential of the photoamplitude phase. Important recent measurements using the RABITT technique include relative delay between argon $3s$ and $3p$ photoemission~\cite{klunder2011probing, guenot2012photoemission}, between emissions from various noble-gas atoms~\cite{guenot2014}, Ar $3p$ photorecombination delay and phase at the $3p$ Cooper minimum (CM)~\cite{schoun2013attosecond}, and the Ar $3p$ emission phase at a Fano autoionizing resonance~\cite{galan2014,kotur2015fanoresonance}.
 
Several theoretical methods~\cite{klunder2011probing, guenot2012photoemission, dahlstrom2012diagrammatic, kheifets2013time, dixit2013, carette2013multiconfigurational, dahlstrom2014, magrakvelidze2015coopermin} have been employed to explain experimental $3s-3p$ relative delay~\cite{klunder2011probing, guenot2012photoemission} in argon with a diverse range of success. Phase and Wigner-Smith delay calculations by us~\cite{magrakvelidze2015coopermin} using the time-dependent local-density approximation (TDLDA) have recently agreed very well with the experiment on argon $3p$ photorecombination around the $3p$ CM~\cite{schoun2013attosecond}. Correlated delays in the emission between atom-fullerene hybrid electrons near Cooper-type minima in the Ar@C$_{60}$ molecule were unraveled using TDLDA~\cite{dixit2013}. For a different variety of spectral minima, which are abundant in the photoemission of cluster systems, TDLDA has probed attosecond structures in the emission delay~\cite{magrakvelidze2015cavitymin}. Other successes of TDLDA include the capture of the full landscape of electronic collective motions in C$_{60}$ driven by photon that produce experimentally detected plasmon resonances~\cite{scully2005,madjet2008}. Very recently, TDLDA has predicted attosecond time delays of the valence photoionizations of C$_{60}$ at the giant plasmon resonance that showed negative delay behavior over a range of the resonance, matching the outcome of a simple semiclassical modeling~\cite{barillot2015}.

The Xe giant dipole resonance (GDR) in the $4d$ ionization continuum has been the focus of myriad theoretical and experimental studies over last decades. This is a unique spectral feature largely originated from a many-body correlation driven collective process involving predominantly 10 inner electrons in the $4d$ subshell~\cite{starace1970, wendin1973, amusia2000, chen2015}. This feature is fundamentally different from non-collective, large resonances, such as, the $3p\rightarrow 3d$ Auger in Mn~\cite{dolmatov2015}. Even recently, with the advent of newer spectroscopic techniques, the Xe GDR continues to remain an eminent testing ground of many-body effects in high-harmonic generation (HHG)~\cite{shiner2011,pabst2013} and free electron laser~\cite{gerken2014}. On the other hand, Xe GDR can also be a particularly interesting spectral laboratory to explore the electronic temporal behavior that can discern the influence of the many-body effect on the emission time. There has only been a single measurement so far of Xe $4d$ delay relative to $5s$ by the attosecond streaking method at about 97.5 eV XUV energy~\cite{verhoef2013}. Furthermore, the study of the anti-resonance spectral features, such as the CM, has also been the subject of long-standing interests both in the older synchrotron-type as well as newer pump-probe RABITT spectroscopies. First, the presence of such minima in pump-probe spectra indicates that the structure of the sample can be temporally probed despite the  presence of an IR pulse during the process. Second, since the ionization strength significantly diminishes at the vicinity of a CM, correlations with other degenerate channels become conspicuous. As a result, CM regions may serve as strategic spectral windows to scrutinize the effects of correlations~\cite{magrakvelidze2015coopermin}. 

In this paper, we present the phase of Xe $4d$ dipole photoionization amplitude and resulting Wigner-Smith time delay calculated in TDLDA; cross sections are presented to only compare TDLDA results with synchrotron measurements. We show how the electron correlation driven collective dynamics at the $4d$ GDR strongly favors an accelerated ionization of electron by producing a negative time delay. Strong temporal variation is also found for the emission at the $4d$ CM as a consequence of the correlation. A succinct description of the method is given in Section II. Sec. III presents the numerical results with discussions. The paper is summarized in Sec. IV. 

\section{A brief description of theory}

Choosing the photon polarization along the $z$-axis, the photoionization dipole transition amplitude in a single-channel independent-particle approximation, which omits electron correlations, is:
\begin{equation}\label{eq1}
d (\mathbf{k}) = \langle \psi_{\mathbf{k}l'} | z | \phi_{nl} \rangle.
\end{equation}
Here $\mathbf{k}$ is the momentum of the continuum electron, $z$ is the one-body dipole operator, $\phi_{nl}$ is the bound wavefunction of the target, and the outgoing spherical continuum wavefunction $\psi_{\mathbf{k}l'}$ is
\begin{equation}\label{eq3}
\psi_{\mathbf{k}l'}(\mathbf{r}) = (8 \pi)^{3/2} \sum_{m} e^{i \eta_{l'}} R_{k l'}(r) 
Y_{l' m}(\mathbf{\Omega_{r}}) Y_{l' m}^{*}(\mathbf{\Omega_{k}})
\end{equation}
with $l'=l\pm1$. In \eq{eq3}, the scattering phase $\eta_{l'}(k)$ contains contributions from both short-range and Coulomb potentials, besides a constant phase $l'\pi/2$, and $R_{k l'}$ is the radial continuum wave.

We calculate the amplitudes $d$ [Eq.\,(\ref{eq1})] using the independent particle LDA method~\cite{madjet2001,stener1997tdldaPI,zangwill1980density}. Here the LDA potential, using the single-particle density $\rho(\mathbf{r}$),
\begin{equation}\label{eq3a}
V_{\scriptsize \mbox{LDA}}(\mathbf{r}) = -\frac{z}{r} + \int d\mathbf{r}'\frac{\rho(\mathbf{r}')}{|\mathbf{r}-\mathbf{r}'|} + V_{\scriptsize \mbox{XC}}[\rho(\mathbf{r})]
\end{equation}
uses the Leeuwen-Baerends (LB) exchange-correlation functional $V_{\scriptsize \mbox{XC}}$~\cite{van1994exchange}, which provides an accurate asymptotic description of the ground state potential. The LDA self-consistently includes an average interaction with the ionic core, and obtains the ground and continuum single-electron properties for various angular momenta in a mean-field approximation. Thus, the LDA is akin to the Hartree-Fock method, albeit an approximation to the (nonlocal) exchange in a local frame. Eq.\,(\ref{eq1}) includes LDA radial matrix elements $\langle R_{kl'} | z | R_{nl} \rangle$. 

The TDLDA, used here to calculate the full transition amplitude, includes many-electron effects and utilizes the advanced Green's function $G$~\cite{zangwill1980density,ekardt1985size,yabana2001}.
In a linear response frame, such as the TDLDA, the photoionization amplitude formally reads
\begin{equation}\label{eq4}
D (\mathbf{k}) = \langle \psi_{\mathbf{k}l'} |\delta V (\mathbf{r}) + z | \phi_{nl} \rangle = d (\mathbf{k}) + \langle \delta V(\mathbf{r}) \rangle,
\end{equation}.
Here $\delta V(\mathbf{r})$ is the complex induced potential that accounts for electron correlations. In TDLDA, $z + \delta V(\mathbf{r})$ is proportional to the induced frequency-dependent changes in the electron density~\cite{madjet2008}. This change is 
\begin{equation}\label{ind-dens}
\delta \rho (\mathbf{r}^{\prime}; \omega) = \int \chi (\mathbf{r}, \mathbf{r}^{\prime}; \omega)
z  d\mathbf{r},
\end{equation}
where the full susceptibility $\chi$ builds the dynamical correlation from the independent-particle LDA susceptibilities
\begin{eqnarray}\label{suscep}
\chi^{0} (\mathbf{r},\mathbf{r}^{\prime };\omega) &=&\sum_{nl}^{occ}\phi _{nl}^{*}
(\mathbf{r})\phi _{nl}(\mathbf{r}^{\prime })\ G(\mathbf{r},\mathbf{r}^{\prime };\epsilon
_{nl}+\omega)  \nonumber \\
&+&\sum_{nl}^{occ}\phi _{nl}(\mathbf{r})\phi _{nl}^{*}(\mathbf{r}^{\prime })\ G^*
(\mathbf{r},\mathbf{r}^{\prime };\epsilon _{nl}-\omega)  
\end{eqnarray}
through the matrix equation $\chi = \chi^0[1-(\partial V/\partial \rho)\chi^0]^{-1}$ involving the variation of the ground-state potential $V$ with respect to the ground-state density $\rho$. The radial components of the full Green's functions in \eq{suscep} are constructed with the regular ($f_L$) and irregular ($g_L$) solutions of the homogeneous radial equation 
\begin{equation}\label{radial-eq}
\left( \frac{1}{r^2} \frac{\partial}{\partial r} r^2 \frac{\partial}{\partial r} 
     - \frac{L(L+1)}{r^2} - V_{\mbox{\scriptsize{LDA}}} 
     + E \right) f_L(g_L) (r;E) = 0
\end{equation}
as
\begin{equation}\label{green}
G_{L}(r,r^{\prime };E)=\frac{2f_{L}(r_{<};E)h_{L}(r_{>};E)}{W [f_{L},h_{L}]}  
\end{equation}
where $W$ represents the Wronskian and $h_{L}= g_{L} + i\; f_{L}$. The TDLDA radial matrix element can thus be written as $\langle R_{kl'} | z + \delta V(r)| R_{nl} \rangle$.

The total $nl$ cross section $\sigma_{nl} = \sum_{l'} \sigma_{nl\rightarrow kl'}$, where $\sigma_{nl\rightarrow kl'}$ is obviously proportional to the modulus square of \eq{eq1} and \eq{eq4}, respectively for LDA and TDLDA, integrated over the photoelectron direction $\Omega_\mathbf{k}$. In a non-angle-resolved measurement, such as RABITT, the total amplitude is directly close to the $\Omega_\mathbf{k}$-integrated dipole matrix elements. Thus, following Ref.~\cite{magrakvelidze2015coopermin}, the TDLDA $nl$ ionization phase is approximated by
\begin{eqnarray}\label{ang-integrated}
\Gamma_{nl} = \arg \left[\sum_{l'}\sqrt{\sigma_{nl\rightarrow kl'}} \exp(i \Gamma_{nl\rightarrow kl'})\right],
\end{eqnarray}
where $\Gamma_{nl\rightarrow kl'}$ are channel phases. The energy differential of $\Gamma_{nl}$ is the Wigner-Smith time delay of the emission from the $nl$ subshell~\cite{wigner1955lower,smith1960lifetime}. \eq{ang-integrated} and the delay derived from it recently well explained the measured 3$p$ photorecombination RABITT measurements~\cite{magrakvelidze2015coopermin}. 

\begin{figure}[h!]
\includegraphics[width=9cm]{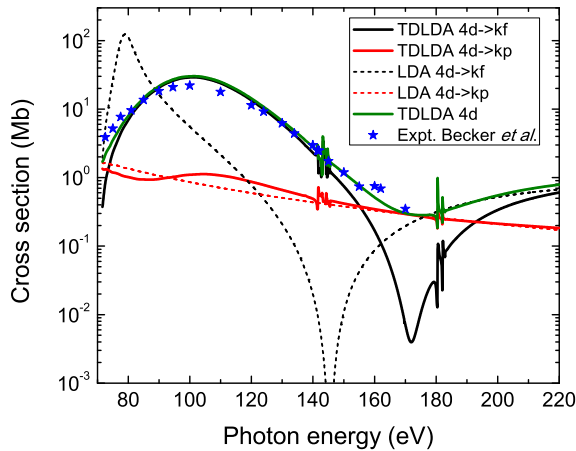}
\caption{(Color online) LDA and TDLDA channel cross sections of the photoionization of Xe $4d$. The TDLDA $4d$ total cross section is compared with the experiment~\cite{becker1989}.} \label{fig1}
\end{figure}

\section{Results and discussions}

Figure 1 presents the LDA cross sections for the two dipole photoionization channels $4d\rightarrow kp,f$, of which the result corresponding to the $kf$ continuum displays a shape resonance and a CM at approximately 80 eV and 145 eV, respectively. The origin of this shape resonance is the transient binding of the photoelectron due to the strong centrifugal barrier of the $f$ continuum~\cite{cooperGDR1964} and the CM is a real zero corresponding to the sign reversal of the radial LDA amplitude that is real~\cite{cooperCM1962}. However, as the correlation is included, that is by TDLDA, the $4d\rightarrow kp$ result is seen to remain largely unchanged, but that of $4d\rightarrow kf$ non-trivially modifies: (i) the shape resonance blue-shifts to 101 eV and broadens, primarily, as a consequence of the collectivization of ten $4d$ electrons via the $4d$ {\em intrachannel} coupling~\cite{amusia2000,brechignac1994}. This resonance is known as the giant dipole resonance (GDR). (ii) The correlation in TDLDA also increases the energy of CM to 172 eV, but this minimum is now not a real zero as the imaginary component of complex $\delta V$ [\eq{eq4}] being nonvanishing contributes some strength at this energy. The total $4d$ TDLDA cross section is also presented in Fig.\,1 and compared with the experiment~\cite{becker1989}, showing very nice agreements. 

\begin{figure}[h!]
\includegraphics[width=9cm]{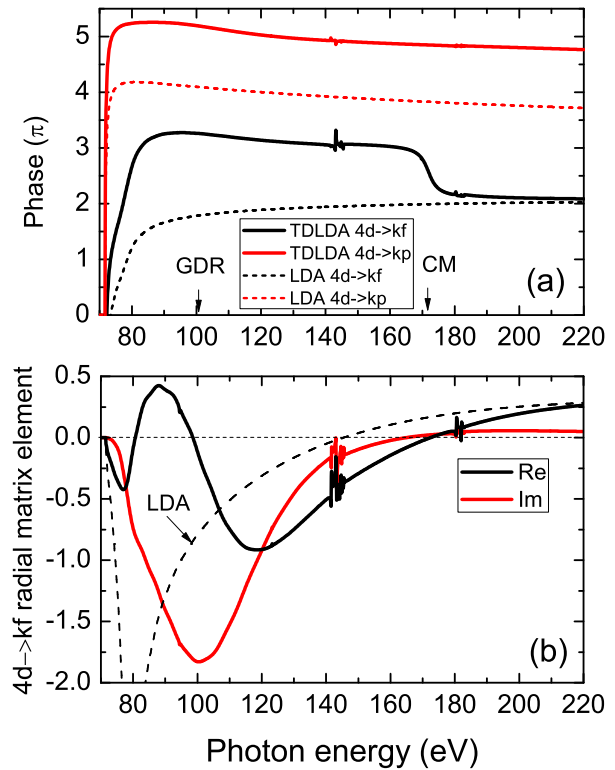}
\caption{(Color online) (a) LDA scattering phases ($\eta$) and TDLDA phases ($\Gamma$) for the Xe $4d$ photoemission channels. The positions of the giant dipole resonance (GDR) at 101 eV and the Cooper minimum (CM) at 172 eV are indicated. (b) Real and imaginary components of the $4d\rightarrow kf$ radial matrix element in TDLDA. The real LDA radial matrix element is also shown.} \label{fig2}
\end{figure}

We now address the phases of the 4d channels whose LDA results, the LDA scattering phases only [$\eta$ in \eq{eq3}], are shown in Figure 2(a). While the lowest energy part of these results is dominated by the long-range Coulomb phase, the short-range phase takes over at higher energies. The difference between LDA and the corresponding TDLDA results ($\Gamma$) in Fig.\,2 uncovers the effect of correlations. This difference is only a constant magnitude shift for the $4d\rightarrow kp$ channel, as seen. In contrast, the effect of correlation is rather dramatic for $4d\rightarrow kf$: The correlation switches the energy gradient of the phase from positive in LDA to negative in TDLDA over the GDR energy-window and leaves its imprint across the CM by inducing a dramatic downshift of about $\pi$ radian. A variation similar to the latter but for the Ar $3p\rightarrow kd$ TDLDA phase near the $3p$ CM was discussed earlier~\cite{magrakvelidze2015coopermin}. This effect transpires from the real part of the TDLDA radial matrix element sloshing through a zero at CM when the imaginary part is nonzero, as seen in Figure 2(b). It may be noted that even though the real LDA radial matrix element also crosses a zero at CM, the resulting discontinuous phase variation and the indeterminate $\arctan (0/0)$ right at CM renders the phase of the LDA radial matrix element unphysical, and hence is ignored. 

Over the GDR region, the interplay between the real and imaginary parts of the $4d\rightarrow kf$ matrix element in Fig.\,2(b) further uncovers the detailed role of correlations. As a function of energy the real part is seen to turn around and sluice through zeros twice at around 80 eV and 101 eV. As characteristics of the collective resonant motions~\cite{wendin1973}, the imaginary component also shows minima at these energies; qualitatively similar results were also recently obtained in the time-dependent configuration interaction singles (TDCIS) calculations~\cite{chen2015}. The lower energy minimum is found significantly weak in TDLDA and its position coincides with the deep minimum of the LDA matrix element (also shown) at the LDA shape resonance (Fig.\,1). This weak structure however shows up only as a very feeble (backward) knee klinged on the broad, dominating minimum in the imaginary part of the TDLDA matrix element at 100 eV. Obviously, at the level of the TDLDA cross section (Fig.\,1), which involves squared modulus of the matrix element, the effect of the low-energy peak is undecipherable. 

\begin{figure}[h!]
\includegraphics[width=9cm]{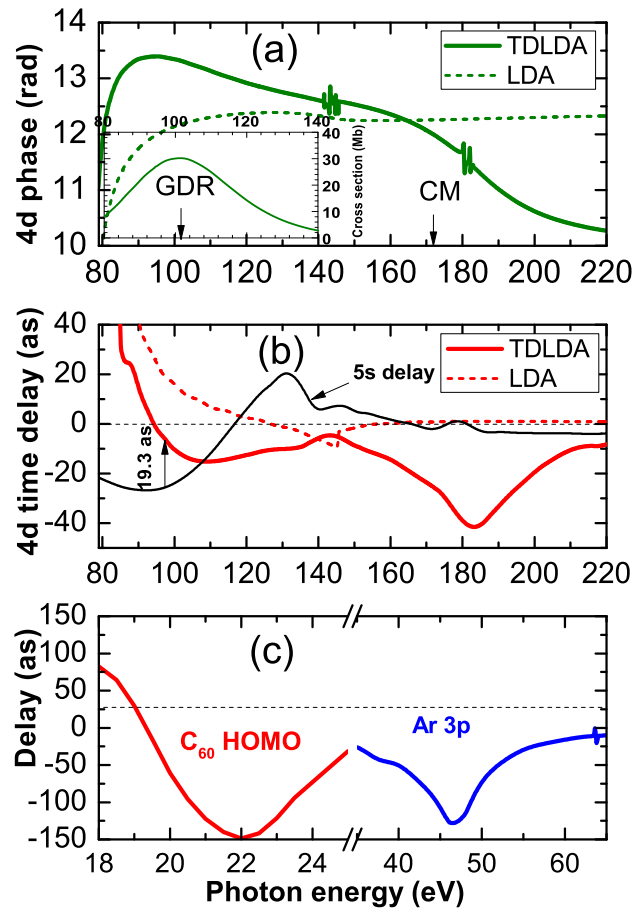}
\caption{(Color online) (a) LDA and TDLDA $4d$ photoionization phase of Xe. The inset carries the $4d$ TDLDA cross section in the GDR region. (b) The corresponding Wigner-Smith time delays. The TDLDA delay for the $5s\rightarrow kp$ channel is also shown. The vertical arrow at 97.5 eV photon-energy indicates 19.3-attosecond TDLDA delay of the $4d$ emission relative to $5s$ which closely agrees with the relative delay of 18 attoseconds observed in streaking measurements~\cite{verhoef2013}. (c) Published emission delays of C$_{60}$ HOMO electron at the plasmon resonance~\cite{barillot2015} and Ar $3p$ electron near $3p$ CM~\cite{magrakvelidze2015coopermin}, both in TDLDA, are shown for the comparison.} \label{fig4}
\end{figure}

Figure 3(a) presents the total $4d$ phase, in LDA [using $\eta$ for $\Gamma$ in \eq{ang-integrated}] and in TDLDA [using \eq{ang-integrated}] where the stronger $4d\rightarrow kf$ channel dominates. Barring the low energy Coulomb region, the LDA phase is quite flat, producing generally small delay in Figure 3(b); a rather innocuous structure at 145 eV is the effect of LDA CM in $\sigma$ in \eq{ang-integrated}. The TDLDA phase shows significant variations both over the GDR and CM regions. In fact, the TDLDA curve produces a broad region of negative slope, mostly over the waning part of GDR (see inset) where the effect of Coulomb repulsion weakens, resulting in a negative delay with a minimum of roughly -16 attoseconds as Fig.\,3(b) delineates. Similar negative delay times were also predicted in TDLDA for HOMO and HOMO-1 photoemission in C$_{60}$ at the giant plasmon resonance energies~\cite{barillot2015}, which are reproduced in Fig.\,3(c). This generic delay behavior at collective resonances is not surprising. When a collective excitation decays through the ionization continuum, the mechanism of the photo-liberation of the electrons become efficient, facilitating a rather faster emission (negative delay). A possible interpretation of the Wigner-Smith delay is the excess time, positive or negative, spent by the electron to reach the continuum in addition to the time it would take in the absence of interactions between the continuum electron and the target. Hence, GDR induced negative delay can be construed as if the emerging electrons feel a transient repulsion from its many body interaction with the residual core that supports resonant collective motions at that particular energy. Due to the richness of many-body physics at the energies of collective response, experiments with the temporal access into Xe $4d$ giant resonance are highly desirable.

Over a wide range surrounding the $4d$ CM, TDLDA also predicts negative delays with a maximum of about $-$43 attoseconds at 185-eV photon energy. The result points to the considerable impact of the electron correlation via the interchannel coupling~\cite{magrakvelidze2015coopermin} of significantly-weakened $4d$ channel with other degenerate open channels around CM. The general shape of the $4d$ phase and delay in Figs.\,3 over this range agrees with those measured at Ar $3p$ CM in the photorecombination process using the RABITT technique~\cite{schoun2013attosecond} which agreed well with our TDLDA calculations~\cite{magrakvelidze2015coopermin}; Ar $3p$ TDLDA delay is shown in Fig.\,3(c) to aid comparisons. Hence, this spectral region can also be attractive for RABITT type experiments to probe the details of electron correlation.  

We also present the Xe $5s$ emission time delay result in Fig.\,3(b). Since the interchannel coupling of weaker $5s\rightarrow kp$ channel with stronger $4d$ channels clones a shape resonance in $5s$ emission as well, negative time delays in $5s$ emission induced by GDR is noted. We particularly point out that our TDLDA $4d-5s$ relative delay of 19.3 attosecond at 97.5 eV closely agrees with 18-attosecond delay measured by the streaking spectroscopy~\cite{verhoef2013}.  

\section{Conclusions}      
In conclusion, a theoretical study of Xe $4d$ photoionization spectral phases and associated Wigner-Smith time delays has been carried out within our TDLDA methodology. Strong spectral variations in the quantum phase of Xe $4d$ emission in both the $4d$ giant resonance and Cooper minimum ranges are noted. The Wigner-Smith time delay derived from this phase indicates a negative delay suggesting faster emission at the resonance. This is likely a ramification of efficient photoelectron emissions driven by electronic collective dynamics. The result corroborates the earlier TDLDA predictions of negative delays for the valence photoionization at the giant plasmon in C$_{60}$~\cite{barillot2015}. Furthermore, the phase and delay of $4d$ electron over the $4d$ Cooper minimum suggest structures very similar to Ar $3p$ emission at its $3p$ minimum which was measured~\cite{schoun2013attosecond} and computed~\cite{magrakvelidze2015coopermin} before. TDLDA produces a reasonable comparison with the time delay detected by the streaking method for Xe $4d$ relative to the Xe $5s$ emission~\cite{verhoef2013}. We hope that the current results will motivate experiments, particularly based on RABITT-type interferometric techniques, to access temporal effects of many-body correlations near the spectrally attractive giant resonance region as well as near the correlation-sensitive Cooper minimum anti-resonances.

\begin{acknowledgments}
The research is supported by the National Science Foundation, USA.
\end{acknowledgments}

\end{document}